\documentclass[graybox]{svmult}
\usepackage{mathptmx}       
\usepackage{helvet}         
\usepackage{courier}        
\usepackage{type1cm}        
\usepackage{graphicx}        
\usepackage{multicol}        
\usepackage[bottom]{footmisc}

\usepackage{subfigure}
\usepackage{amssymb}
\usepackage{amsmath}
\usepackage{amsfonts}

\begin{document}
 
\title*{Propagation of compressional elastic waves through a 1-D medium with contact nonlinearities}
\titlerunning{Scattering of elastic waves by nonlinear contacts}




\author{B.~Lombard$^{(1)}$ \and J.~Piraux$^{(1)}$}


\institute{(1) LMA - CNRS, 31 chemin Joseph Aiguier - 13402 Marseille, France.}

\maketitle

\abstract{Propagation of monochromatic elastic waves across cracks is investigated in 1D, both theoretically and numerically. Cracks are modeled by nonlinear jump conditions. The mean dilatation of a single crack and the generation of harmonics are estimated by a perturbation analysis, and computed by the harmonic balance method. With a periodic and finite network of cracks, direct numerical simulations are performed and compared with Bloch-Floquet's analysis.}



\section{Introduction}\label{SecIntroduction}

Failure processes resulting in a crack generally produce rough crack faces. Once crack opening has taken place, and the crack faces have undergone slight relative sliding displacement, the crack will never completely close again due to the nonconforming surfaces in partial contact. A complicated interaction between crack faces is expected, depending strongly on the magnitudes of the tractions transmitted across the rough surfaces in contact \cite{PECORARI03}.

The interaction of ultrasonic waves with cracks has been investigated by many authors, assuming that the wavelength is much larger than a characteristic length of the roughness of the contacting surfaces. Linear slip-displacement models of crack-face interaction have been widely used \cite{SCHOENBERG80,PYRAK90}. However, a non-physical penetration of contacting surfaces may occur in linear models. Moreover, laboratory experiments have shown that methods of non-destructive evaluation based on linear models may fail to detect partially closed cracks \cite{SOLODOV98}.

Here, we study wave propagation with a nonlinear model of contact proposed in \cite{NORRIS82,BANDIS83}. A monochromatic compressional wave propagates normally to a plane flaw surface, leading to a 1D problem detailed in section \ref{SecModel}. Analysis of scattered fields is performed in section \ref{SecAnalysis}. With a single crack, the generation of harmonics and the mean dilatation of the crack are addressed analytically and numerically. Propagation through periodic networks of contact nonlinearities is studied by Bloch-Floquet's analysis \cite{NAKAGAWA00} and simulations. Numerical experiments are proposed in section \ref{SecNum}. Conclusions are drawn and future perspectives are suggested in section \ref{SecConclusion}.

\section{Problem statement}\label{SecModel}

\subsection{Configuration}\label{SecConfig}

\begin{figure}[htbp]
\begin{center}
\includegraphics[scale=0.7]{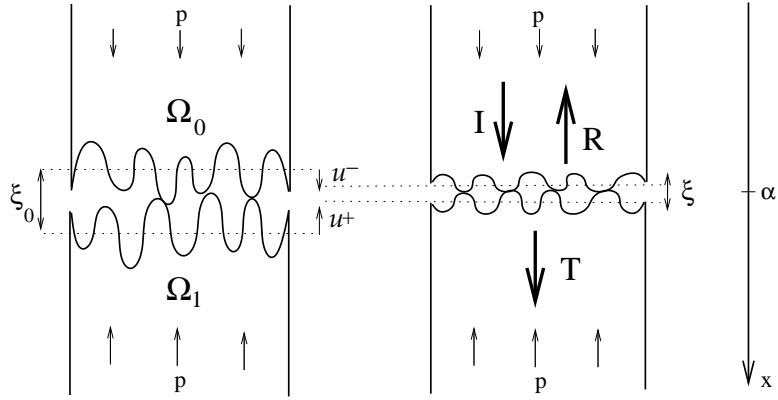}
\caption{Elastic media $\Omega_0$ and $\Omega_1$ separated by vacuum, with rough contact surfaces. Static (left) and dynamic (right) case, with incident (I), reflected (R) and transmitted (T) waves.}
\label{FigCrack}
\end{center}
\end{figure}

We consider a single crack with rough faces separating two media $\Omega_0$ and $\Omega_1$ linearly elastic and isotropic, with density $\rho$ and elastic speed of the compressional waves $c$. These parameters are piecewise constant and may be discontinuous around the crack: $(\rho_0,\,c_0)$ if $x \in \Omega_0$, $(\rho_1,\,c_1)$ if $x \in \Omega_1.$ The media are subject to a constant static stress $p$. At rest, the distance between planes of average height is $\xi_0(p)>0$ (figure \ref{FigCrack}, left). An incident monochromatic wave, emitted by a ponctual stress source at $x=x_s$ in $\Omega_0$, gives rise to reflected (in $\Omega_0$) and transmitted (in $\Omega_1$) compressional waves. These perturbations in $\Omega_0$ and $\Omega_1$ are described by the 1D elastodynamic equations 
\begin{equation}
\rho \,\frac{\textstyle \partial \,v}{\textstyle \partial \,t}=\frac{\textstyle \partial\,\sigma}{\textstyle \,\partial\, x},\qquad
\frac{\textstyle \partial \,\sigma}{\textstyle \partial \,t}=
\rho \,c^2 \,\frac{\textstyle \partial \,v}{\textstyle \partial \,x}+2\,\rho_0\,c_0^2\,v_0\,\delta(x-x_s)\,\sin \omega t,
\label{LCscal}
\end{equation}
where $v_0$ is the amplitude of the incident elastic velocity, and $\omega=2\,\pi\,f$ is the angular frequency of the source. The elastic velocity $v=\frac{\partial\,u}{\partial\,t}$, the elastic displacement $u$, and the elastic stress perturbation $\sigma$ around $p$, are averaged fields per unit area in crack's plane. The dynamic stresses induced by the elastic waves affect the thickness $\xi(t)$ of the crack (figure \ref{FigCrack}, right). The constraint
\begin{equation}
\xi=\xi_0+[u]\geq \xi_0-\delta>0
\label{Penetration}
\end{equation}
must be satisfied, where $[u]=u^+-u^-$ is the difference between the elastic displacements on the two sides of the crack, and $\delta(p)>0$ is the {\it maximum allowable closure} \cite{BANDIS83}. We also assume that the wavelengths are much larger than $\xi$, neglecting the propagation time through the crack, and replacing it by a zero-thickness {\it interface} at $x=\alpha$: $[u]=[u(\alpha,\,t)]=u(\alpha^+,\,t)-u(\alpha^-,\,t)$. 

\subsection{Contact law}\label{SecContact}

\begin{figure}[htbp]
\begin{center}
\includegraphics[scale=0.7]{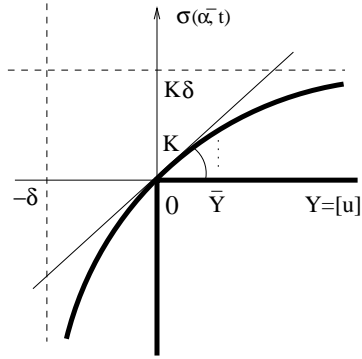}
\end{center}
\caption{Nonlinear relation between stress and slip displacement.}
\label{FigBB}
\end{figure}

Cracks are classically modeled by linear jump conditions \cite{SCHOENBERG80} with {\it stiffness} $K$:
\begin{equation}
\left[\sigma(\alpha,\,t)\right]= 0,\qquad
\left[u(\alpha,\,t)\right]=\frac{\textstyle 1}{\textstyle K}\,\sigma(\alpha^\pm,\,t).
\label{JClin}
\end{equation}
If $K \rightarrow +\infty$, welded conditions are recovered. Conditions (\ref{JClin}) violate (\ref{Penetration}) under large compression loadings: $\sigma(\alpha^\pm,\,t)<-K\,\delta \, \Rightarrow \, \xi <\xi_0-\delta$. Hence, the linear regime induced by (\ref{JClin}) is realistic only with very small perturbations. With larger ones, nonlinear jump conditions are required.

Compression loading increases the number and the surface of contacting faces. Consequently, a smaller stress is needed to open than to close a crack of a given displacement; an infinite stress is even required to close the crack faces completely. This behavior can be modeled by the global jump conditions proposed in \cite{NORRIS82,BANDIS83}  
\begin{equation}
\left[\sigma(\alpha,\,t)\right]= 0,\qquad
\left[u(\alpha,\,t)\right]= \frac{\textstyle 1}{\textstyle K}\,\frac{\textstyle \sigma(\alpha^\pm, \,t)}{\textstyle \displaystyle 1-\sigma(\alpha^\pm,\,t)/(K\,\delta)},
\label{JCBB}
\end{equation}
satisfying (\ref{Penetration}) and implying $\sigma(\alpha^\pm,\,t)<K\,\delta$. The second relation in (\ref{JCBB}) is sketched in figure \ref{FigBB}. The straight line with a slope $K$ tangential to the hyperbola at the origin amounts to the linear jump conditions (\ref{JClin}); as deduced from (\ref{JCBB}), the linear regime is valid only if $|\sigma(\alpha^\pm,\,t)| \ll K\,\delta$. A second limit-case not investigated here is obtained if $\delta\rightarrow 0$ and $K$ bounded: the hyperbola tends towards the nondifferentiable graph of the unilateral contact, denoted by bold straight segments in figure \ref{FigBB}. 

\section{Analysis of scattered fields}\label{SecAnalysis}

\subsection{Single crack}\label{SubSecSingle}

{\bf Analytical approach}. The scattered fields can be expressed in terms of $Y(t)=[u(\alpha,\,t)]$. Following \cite{RICHARDSON79,BIWA04} gives the nonlinear ordinary differential equation (ODE) 
\begin{equation}
\frac{\textstyle d\,Y}{\textstyle d\,t}+\beta\,\frac{\textstyle Y}{\textstyle 1+Y/\delta}=2\,v_0\,\sin \omega\,t,
\label{PA_ODE_Y}
\end{equation}
satisfied by $Y$, with $\beta=K((\rho_0\,c_0)^{-1}+(\rho_1\,c_1)^{-1})$. Inspection of (\ref{PA_ODE_Y}) and dimensional analysis show that $Y$ is $\delta$ times a function of $\frac{v_0}{\beta\,\delta}$, $\frac{\omega}{\beta}$ and $\omega \,t$. To solve (\ref{PA_ODE_Y}), we assume $|Y|/\delta<1$, which leads to the series 
\begin{equation}
\frac{\textstyle d\,Y}{\textstyle d\,t}+\beta\,\sum_{n=0}^\infty\frac{\textstyle (-1)^n}{\textstyle \delta^n}\,Y^{n+1}=2\,v_0\,\sin \omega\,t.
\label{PA_Taylor}
\end{equation}
An approximate solution of (\ref{PA_Taylor}) is sought by a {\it perturbation approximation} (PA) \cite{BENDER}
\begin{equation}
Y(t)=\displaystyle \sum_{n=0}^{\infty}Y_n(t)
=\displaystyle \sum_{n=0}^{\infty}\frac{\textstyle 1}{\textstyle \delta^n}\,y_n(t).
\label{PA_Y}
\end{equation}
Nothing ensures that (\ref{PA_Y}) converges: $|Y|/\delta<1$ is always satisfied when $Y<0$, but it is true when $Y>0$ only if $\frac{v_0}{\beta\,\delta}$ is sufficiently small. Plugging (\ref{PA_Y}) into (\ref{PA_Taylor}) and identifying the terms with identical power of $\delta$ leads to an infinite series
\begin{equation}
\begin{array}{l}
\displaystyle
\frac{\textstyle d\,y_0}{\textstyle d\,t}+\beta\,y_0=2\,v_0\,\sin \omega\,t,\\
[8pt]
\displaystyle
\frac{\textstyle d\,y_1}{\textstyle d\,t}+\beta\, y_1=\displaystyle \beta\,y_0^2,\\
[8pt]
\displaystyle
\frac{\textstyle d\,y_2}{\textstyle d\,t}+\beta\, y_2=\beta\,y_0\left(2\,y_1-y_0^2\right),
\end{array}
\label{PA_ODE}
\end{equation}
and so on. There is no influence of the truncation order $N$ on the accuracy of the solution $Y_n$ with $n<N$. The recursive and linear ODE are much simpler to solve than the nonlinear ODE (\ref{PA_ODE_Y}), even if computing $Y_n$ with $n\geq 2$ is cumbersome. This computation has been automatized with computer algebra tools. Since the number of terms in $Y_n$ is roughly $2^{2\,n+1}$, very high orders are currently out of reach. Computing $Y_n$ up to $N=8$ takes 20 mn on a Pentium 3 GHz. We detail the case $N=1$. Setting $\varphi_0=\arctan \frac{\omega}{\beta}$, the zero-th order periodic solution of (\ref{PA_Y})-(\ref{PA_ODE}) is
\begin{equation}
Y_0(t)=2\,\frac{\textstyle v_0}{\textstyle \beta}\,\frac{\textstyle 1}{\textstyle \displaystyle \sqrt{\displaystyle 1+\left(\omega/\beta\right)^2}}\,\sin(\omega\,t-\varphi_0).
\label{PA_Y0}
\end{equation}
Setting $\varphi_1=\arctan \frac{\beta}{2\,\omega}$, the first-order periodic solution of (\ref{PA_Y})-(\ref{PA_ODE}) is
\begin{equation} 
Y_1(t)=2\,\frac{\textstyle v_0^2}{\textstyle \beta^2\,\delta}\,\,\frac{\textstyle 1}{\textstyle 1+\displaystyle \left(\omega/\beta\right)^2}\,\left(1-\frac{\textstyle 1}{\textstyle \sqrt{\displaystyle 1+\left(2\,\omega/\beta\right)^2}}\,\sin(2\,\omega\,t-2\,\varphi_0+\varphi_1)\right).
\label{PA_Y1}
\end{equation}
Two properties are deduced from (\ref{PA_Y1}). First, the mean value of $Y_1$ is non-null
\begin{equation}
\overline{Y}_1=2\,\frac{\textstyle v_0^2}{\textstyle \beta^2\,\delta}\,\,\frac{\textstyle 1}{\textstyle 1+\displaystyle \left(\omega/\beta\right)^2}>0.
\label{PA_DC1}
\end{equation}
More generally, the mean value $\overline{Y}_{2n+1}$ of $Y_{2n+1}$ is proportional to $\delta\left(\frac{v_0}{\beta\,\delta}\right)^{2n+2}$, and $\overline{Y}_{2n}=0$. The mean thickness of the crack $\overline{\xi}$ deduced from (\ref{Penetration}) satisfies $\overline{\xi}=\xi_0+\overline{Y}>\xi_0$, with $\overline{Y}=\sum \overline{Y}_n$. The dilatation predicted here is similar to the {\it DC} signal measured by \cite{KORSHAK02} with glass-piezoceramic interface. The second property deduced from (\ref{PA_Y1}) concerns the term with angular frequency $2\,\omega$. Its importance is quantified by the ratio of amplitudes between sinusoidal terms in (\ref{PA_Y1}) and (\ref{PA_Y0}) 
\begin{equation}
\gamma_2=\frac{\textstyle v_0}{\textstyle \beta\,\delta}\,\frac{\textstyle 1}{\textstyle 
\sqrt{\displaystyle 1+\left(\omega/\beta\right)^2}\,\sqrt{\displaystyle 1+\left(2\,\omega/\beta\right)^2}}.
\label{PA_gamma}
\end{equation}
Consequently, nonlinear effects increase with $\frac{v_0}{\beta\,\delta}$ and decrease with $\frac{\omega}{\beta}$.\\

{\bf Numerical approach}. To compute the scattered fields with high accuracy and no limitation about the range of validity, one implements the numerical {\it harmonic balance method} (HBM). The periodic elastic displacements are written ($k_0=\omega/c_0$, $k_1=\omega/c_1$)
\begin{equation}
\begin{array}{l}
\displaystyle
u_I(x,\,t)=\frac{\textstyle v_0}{\textstyle \omega}\,\left\{\cos(\omega\,t-k_0\,x)-1\right\},\\
[6pt]
\displaystyle
u_R(x,\,t)=R_0^a+\sum_{n=1}^\infty\left\{R_n^a\,\sin n\,(\omega\,t+k_0\,x)+R_n^b\,\cos n\,(\omega\,t+k_0\,x)\right\},\\
[6pt]
\displaystyle
u_T(x,\,t)=T_0^a+\sum_{n=1}^\infty\left\{T_n^a\,\sin n\,(\omega\,t-k_1\,x)+T_n^b\,\cos n\,(\omega\,t-k_1\,x)\right\}.
\end{array}
\label{HBM_U}
\end{equation}
The elastic stresses are deduced from (\ref{HBM_U}). Fields are truncated at $N$ and injected in (\ref{JCBB}). The truncation implies that $R_n^{a,b}$ and $T_n^{a,b}$ depend on $N$. The terms with identical trigonometric arguments are put together, and the terms with a trigonometric argument greater than $N\,\omega\,t$ are removed. The first condition (\ref{JCBB}) implies $2\,N$ linear equations, without $R_0^a$ and $T_0^a$. The second condition (\ref{JCBB}) implies $2\,N+1$ nonlinear equations, including $T_0^a-R_0^a+v_0/\omega$. Finally, we get a $(4\,N+1)\times (4\,N+1)$ nonlinear system with first-order or second-order polynomial entries
\begin{equation}
\begin{array}{l}
\boldsymbol{F}(\boldsymbol{X})=\boldsymbol{0},\qquad
\boldsymbol{X}=\left(\overline{Y},\,R_1^a,\,T_1^a,\,R_1^b,\,T_1^b,...,\,R_N^b,\,T_N^b\right)^T,\qquad
\displaystyle
\overline{Y}=T_0^a-R_0^a+\frac{\textstyle v_0}{\textstyle \omega}.
\end{array}
\label{HBM_X}
\end{equation}
To solve (\ref{HBM_X}), three cases may be considered:
\begin{itemize}
\item {\it linear regime, for all $N$}: (\ref{HBM_X}) becomes a linear system whose solution is easy to compute analytically. In this limit-case, $\overline{Y}=0$ and $R_n^{a,b}=T_n^{a,b}=0$ if $n\geq2$;
\item {\it nonlinear case, $N=1$}: (\ref{HBM_X}) can be solved analytically. A detailed study shows that the solution may not be unique: if $v_0$ is lower than a critical value ${\tilde v}_0$, there exists only one real root; if $v_0>{\tilde v}_0$, there are three real roots. Moreover, $\overline{Y}>0$; an approximation of this jump recovers the value $\overline{Y}_1$ deduced from the PA (\ref{PA_DC1});
\item {\it nonlinear case, $N>1$}: (\ref{HBM_X}) is solved numerically by the Newton-Raphson method. The determination of $\boldsymbol{F}$ and of its jacobian $\boldsymbol{J}$ has been automatized with computer algebra tools: computing $\boldsymbol{F}$ and $\boldsymbol{J}$ with $N=100$ roughly takes 1 second on a Pentium 3 GHz. The root of (\ref{HBM_X}) may be not unique, like with $N=1$ and $v_0>{\tilde v}_0$. The initialization must therefore be done carefully, e.g. using the exact values of the first 5 components of $\boldsymbol{X}$ at $N=1$, and setting the other components to zero. This simple initialization works up to $v_0\approx {\tilde v}_0$. With stronger nonlinearities, this approach is coupled with a basic continuation. 
\end{itemize}

\subsection{Network of cracks}\label{SubSecMultiple}

{\bf Analytical approach}. A Bloch-Floquet's analysis \cite{NAKAGAWA00} is applied to an infinite and periodic network in linear regime. With constant parameters, the dispersion relation is
\begin{equation}
\cos {\hat k}h=\cos kh -\frac{\textstyle \rho\,c\,\omega}{\textstyle 2\,K}\,\sin kh,
\label{FloquetDisp}
\end{equation}
where ${\hat k}$ is the effective wavenumber and $h$ is the spacing between cracks. If 
\begin{equation}
-\left|\frac{\textstyle 1-\theta}{\textstyle 1+\theta}\right| \leq \cos kh \leq \left|\frac{\textstyle 1-\theta}{\textstyle 1+\theta}\right|, \qquad
\theta=\left(\frac{\textstyle \rho\,c\,\omega}{\textstyle 2\,K}\right)^2,
\label{FloquetPB}
\end{equation}
the waves are not attenuated. Otherwise, the waves are evanescent with decay
\begin{equation}
\Im\mbox{m}\,{\hat k}=-\frac{\textstyle 1}{\textstyle h}\,\cosh^{-1}\left(\cos kh -\frac{\textstyle \rho\,c\,\omega}{\textstyle 2\,K}\,\sin kh\right).
\label{FloquetAtt}
\end{equation}
{\bf Numerical approach}. The analysis is much harder in nonlinear regime, since successive harmonics generated across cracks may not belong to the same pass-band structure. Consequently, nonlinear regime as well as non-periodic configurations are investigated by {\it direct numerical simulations} (DNS). A fourth-order ADER scheme solves (\ref{LCscal}) in the time domain. The jump conditions (\ref{JCBB}) are enforced numerically by an interface method \cite{LOMBARD07}. With high nonlinearities, space-time mesh refinement is implemented around the cracks to discretize correctly the stiff fronts. Obviously, this approach also works with a single crack, hence it will be used to get reference solutions in each experiment.

\section{Numerical experiments}\label{SecNum}

\begin{figure}[htbp]
\begin{center}
\begin{tabular}{ccc}
(a) & (b)\\
\includegraphics[scale=0.28]{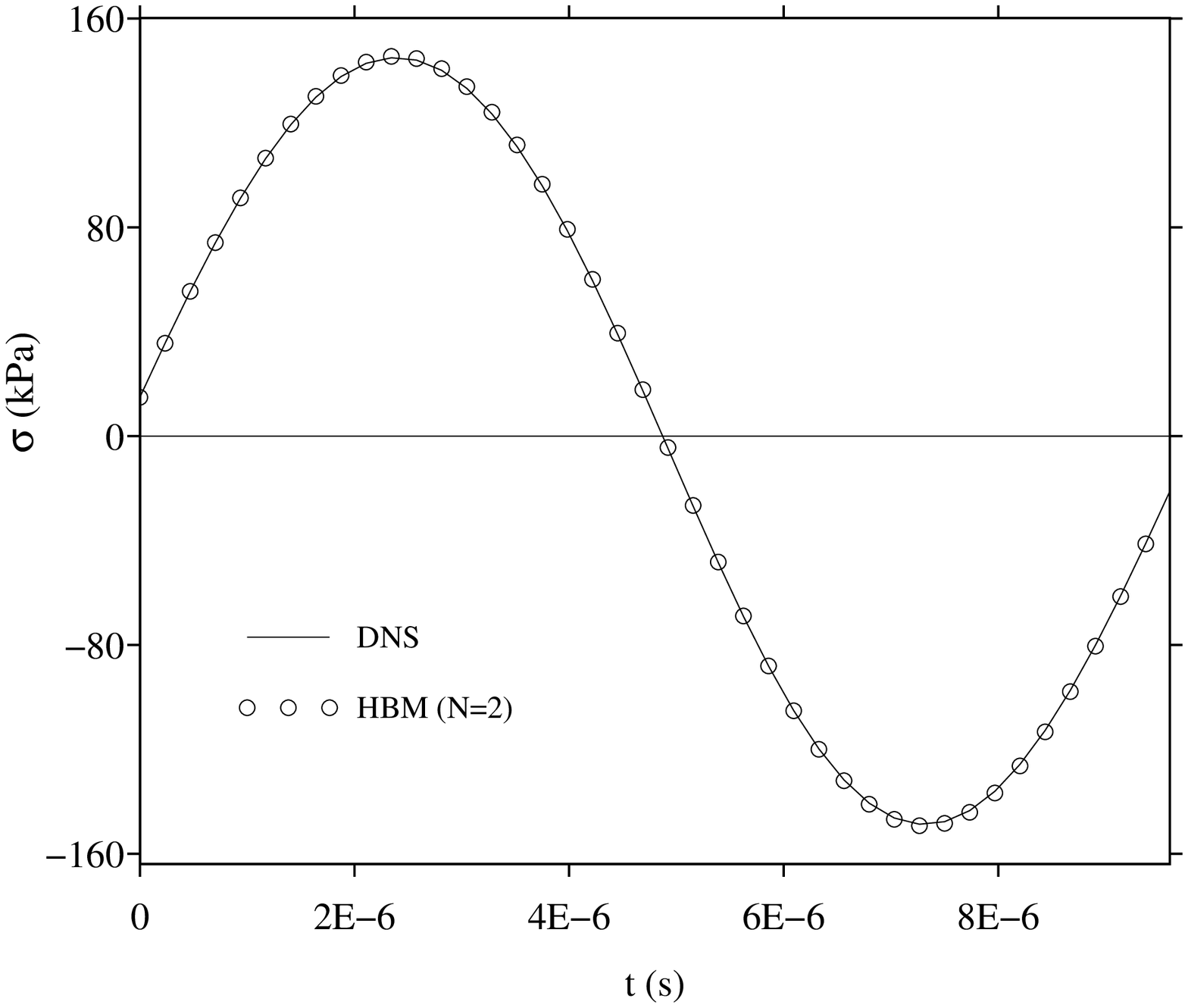}&
\includegraphics[scale=0.28]{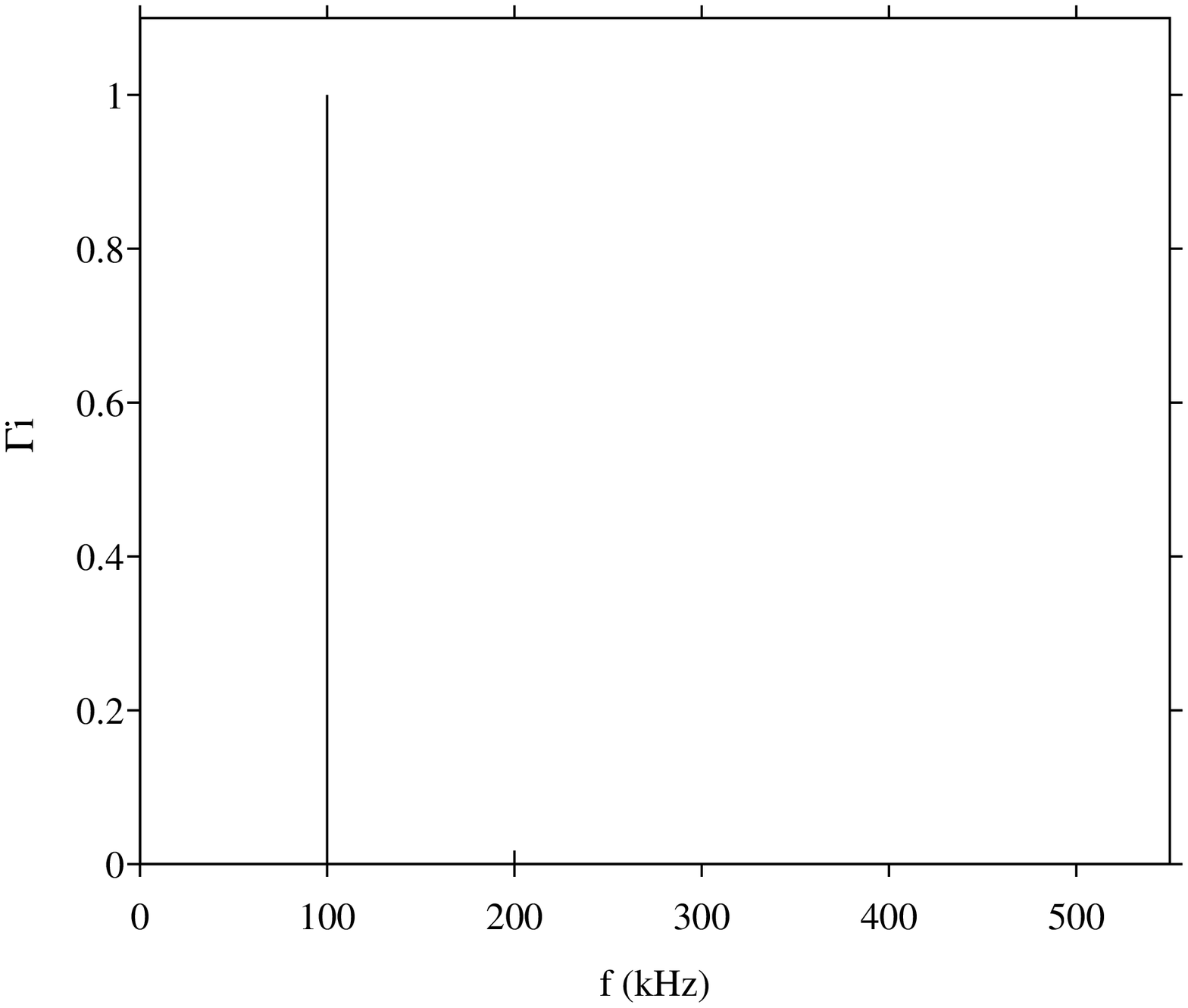}\\
(c) & (d)\\
\includegraphics[scale=0.28]{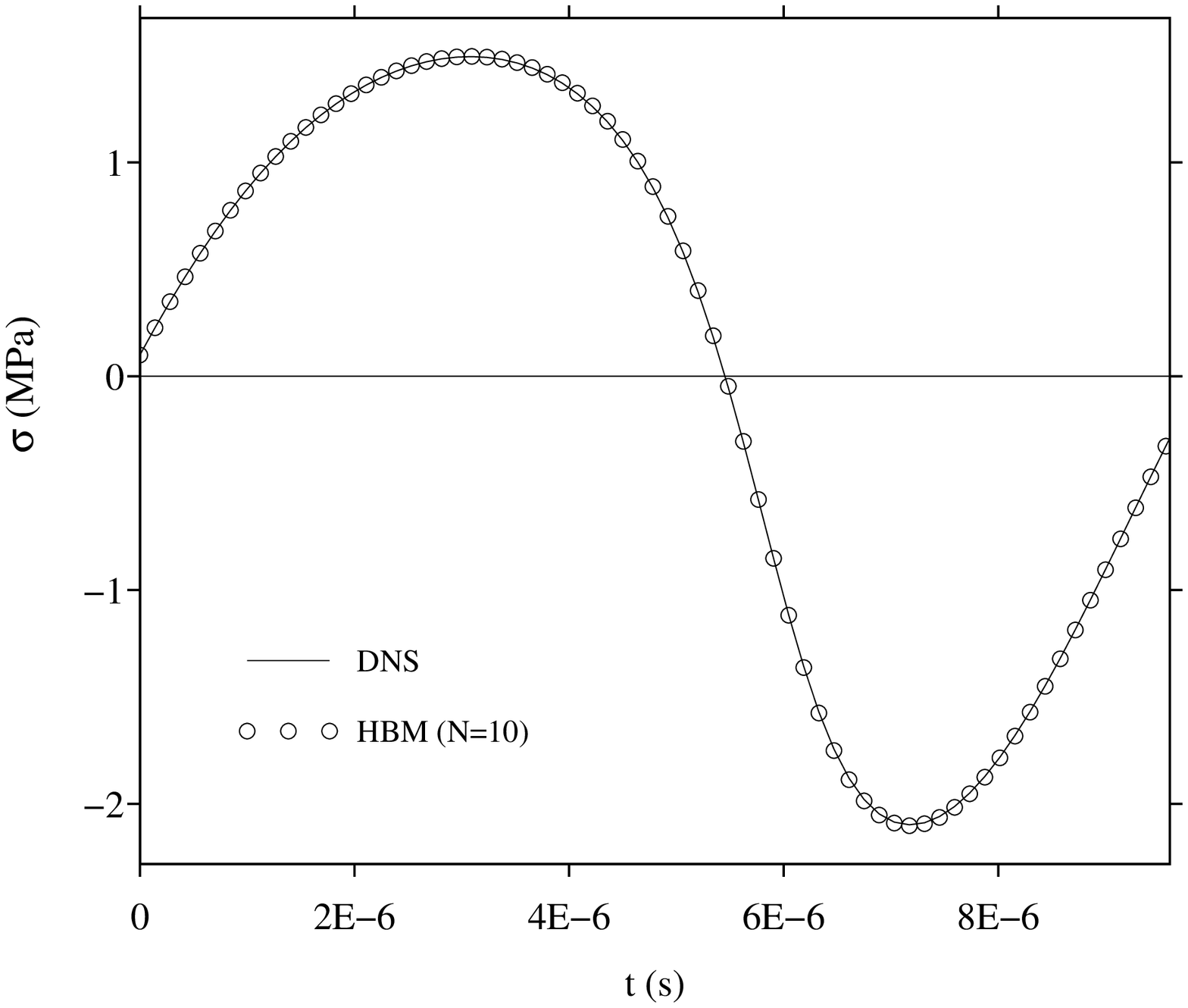}&
\includegraphics[scale=0.28]{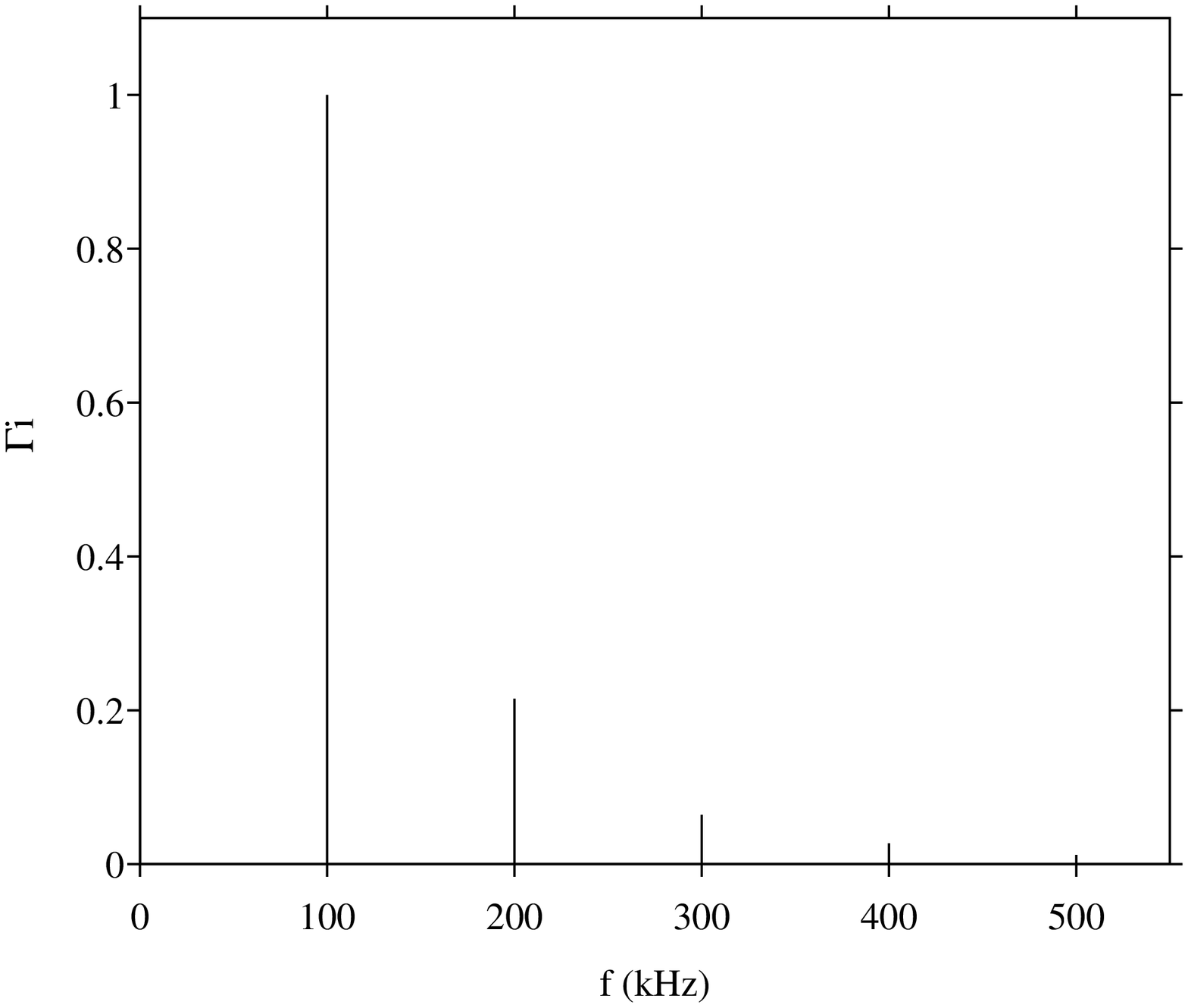}\\
(e) & (f)\\
\includegraphics[scale=0.28]{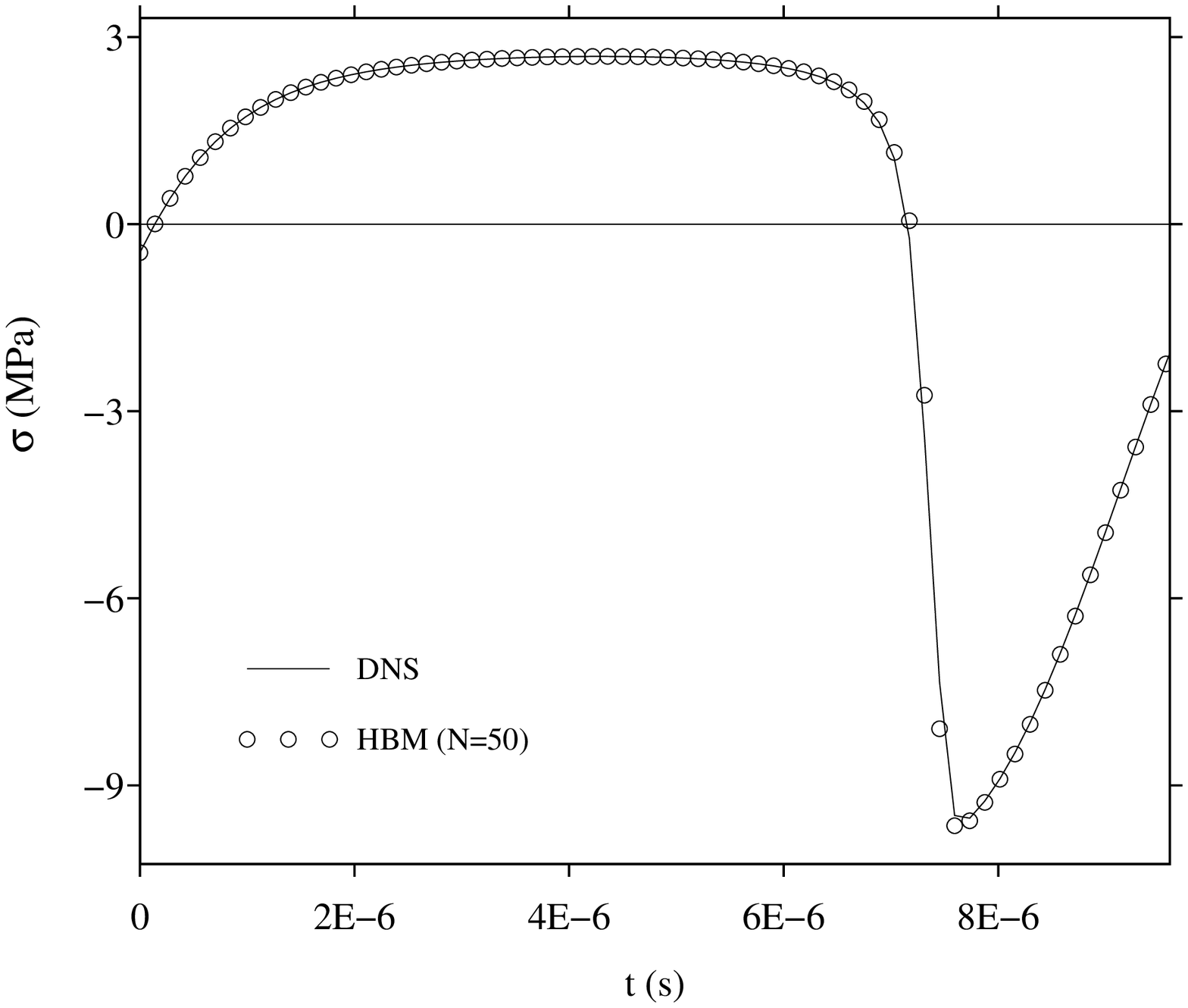}&
\includegraphics[scale=0.28]{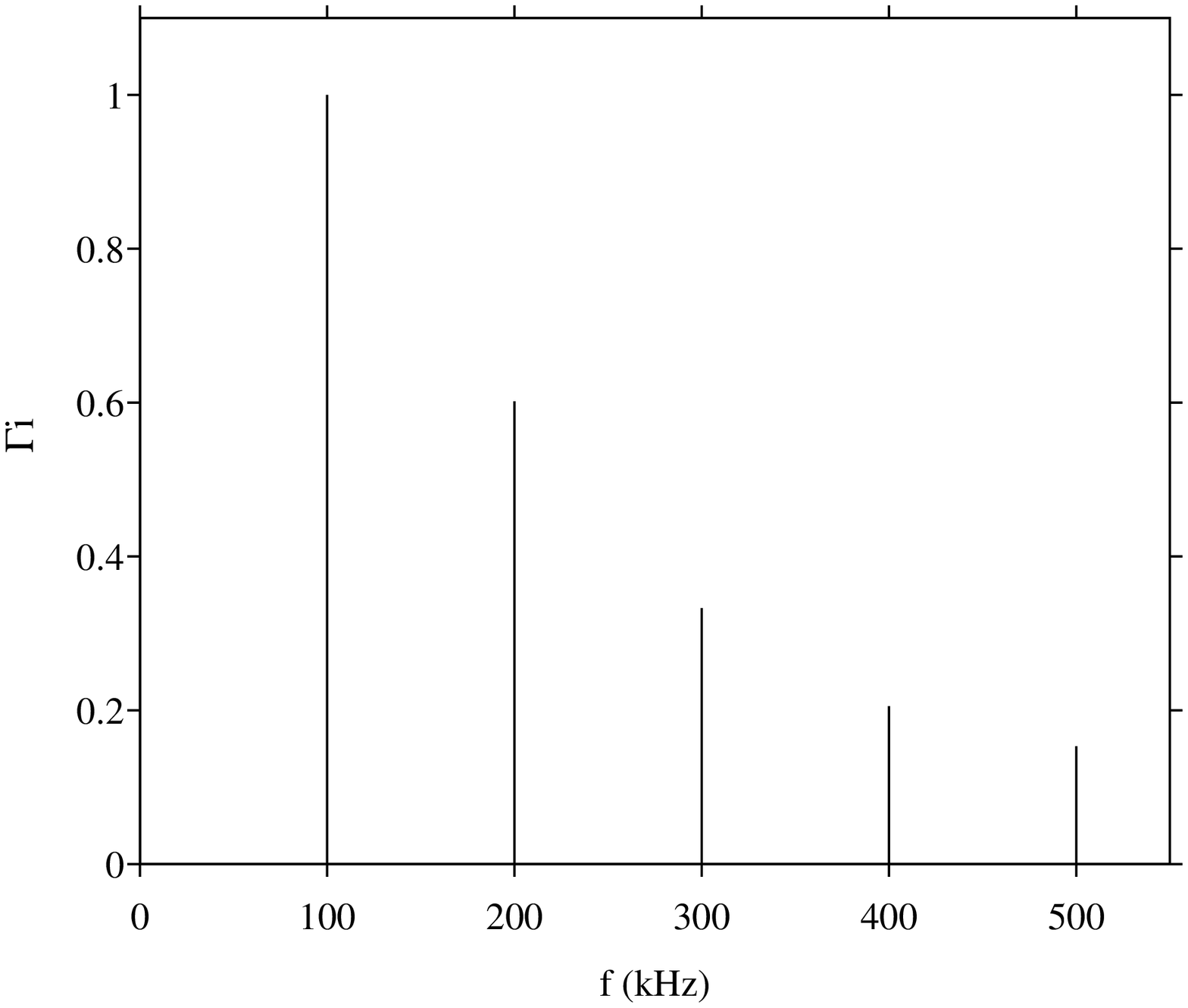}
\end{tabular}
\end{center}
\caption{Transmitted stress, with $v_0=0.01$ m/s (a-b), $v_0=0.13$ m/s (c-d), $v_0=0.6$ m/s (e-f). Left column: values of $\sigma_T$ on a period; right column: normalized harmonics $\Gamma_i$ ($i=1,\,...,\,5$).}
\label{FigTest1}
\end{figure}

A crack at $\alpha=0.5$ m in aluminium is studied: $\rho_0=\rho_1=2600\,\mbox{ kg/m}^3$, $c_0=c_1=6400\,\mbox{ m/s}$, $K=10^{13}$ Pa/m, $\delta=3\,10^{-7}\,\mbox{m}$. Three amplitudes $v_0$ are considered: $0.01$ m/s, $0.13$ m/s, and $0.6$ m/s. The source is at $x_s=0.25$ m, and $f=100$ kHz.

\begin{figure}[htbp]
\begin{center}
\begin{tabular}{cc}
(a) & (b)\\
\includegraphics[scale=0.28]{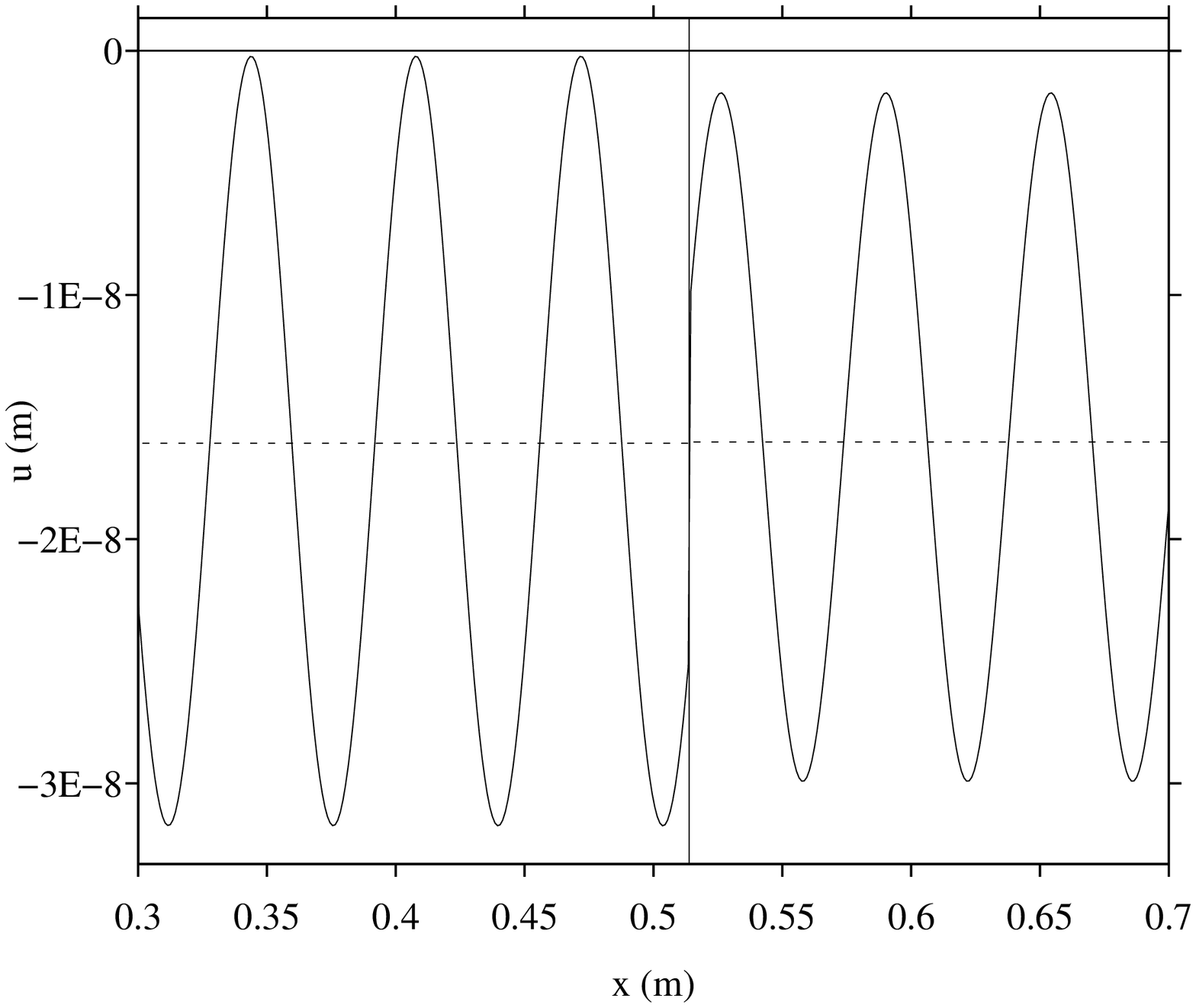}&
\includegraphics[scale=0.28]{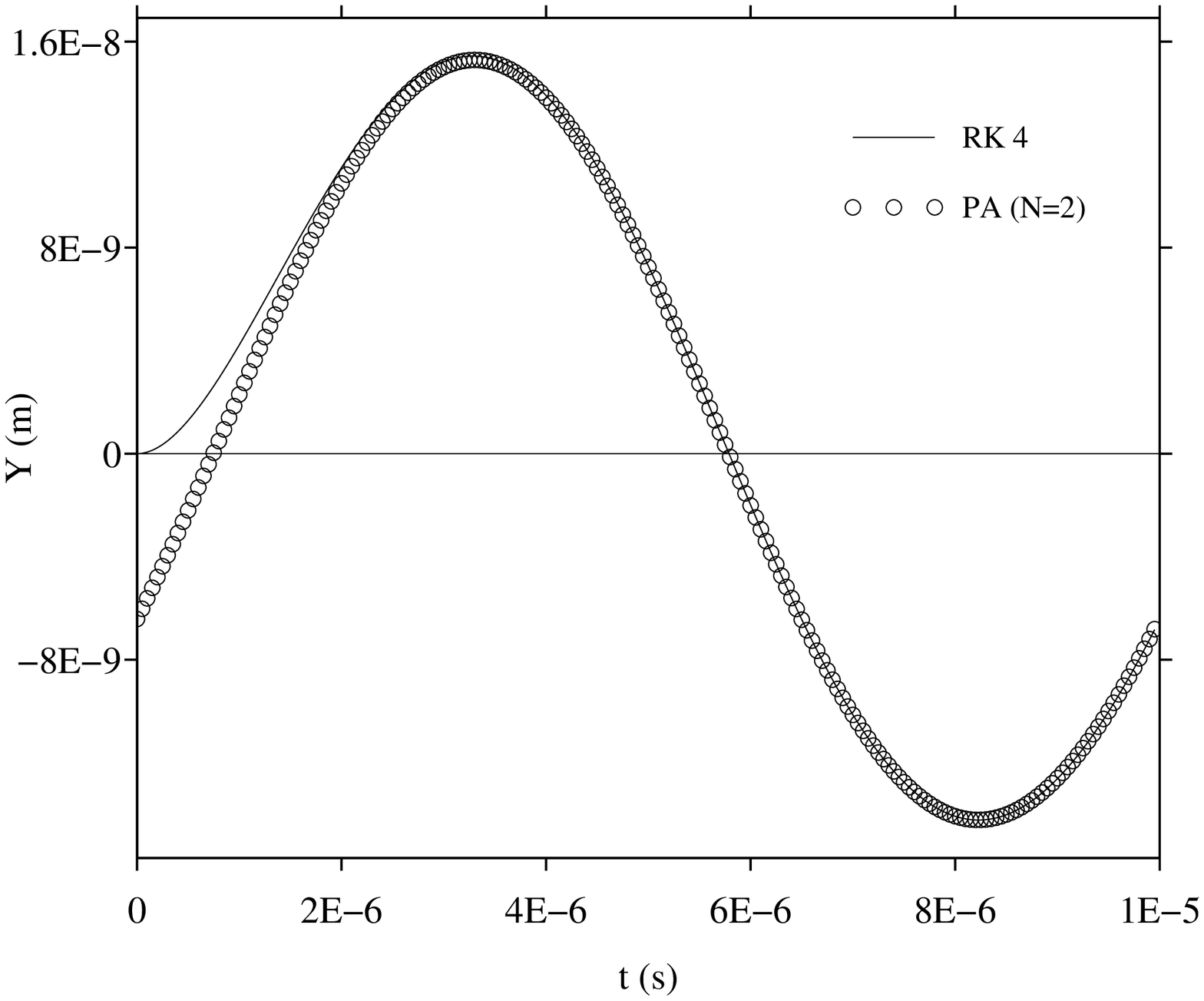}\\
(c) & (d)\\
\includegraphics[scale=0.28]{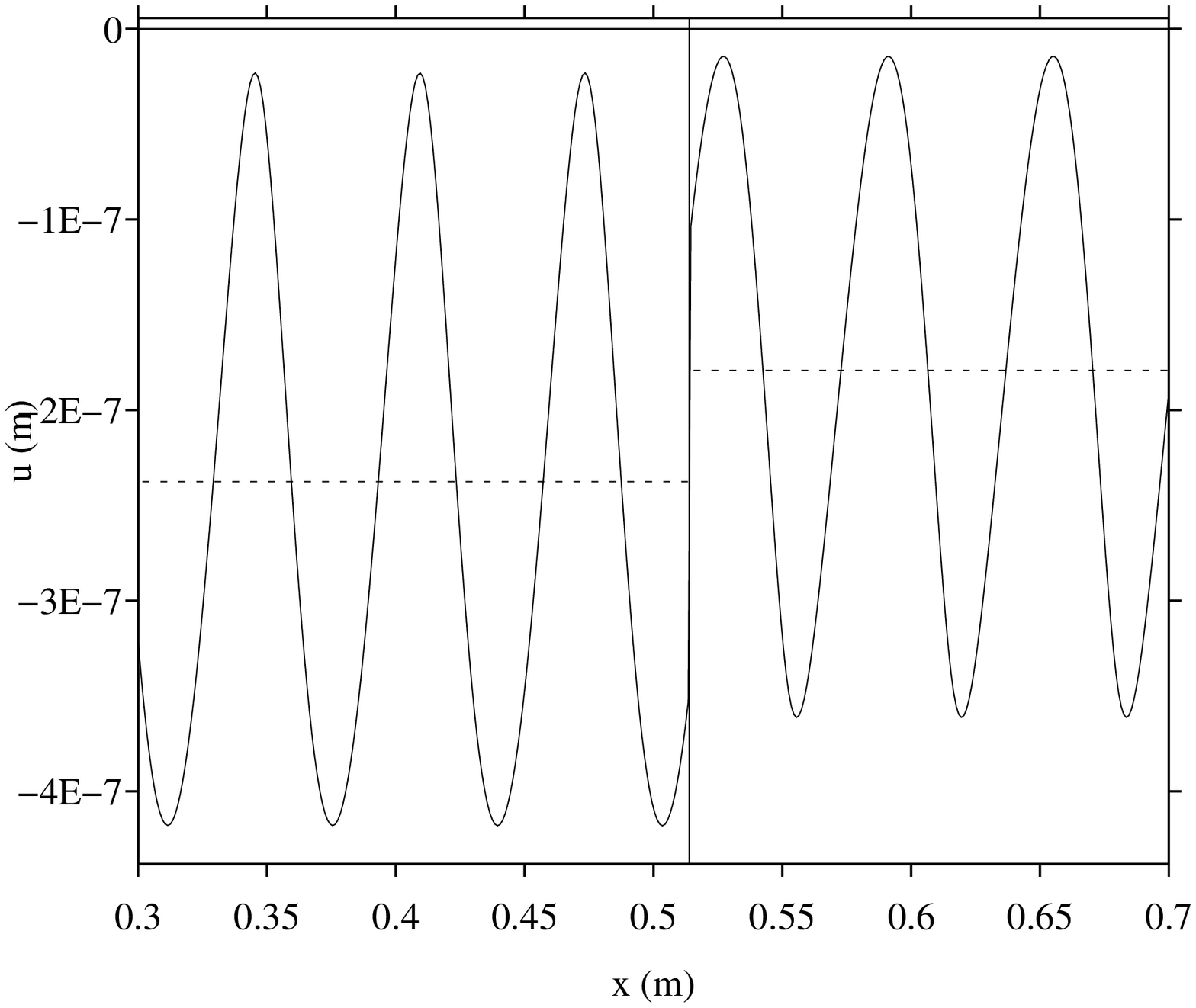}&
\includegraphics[scale=0.28]{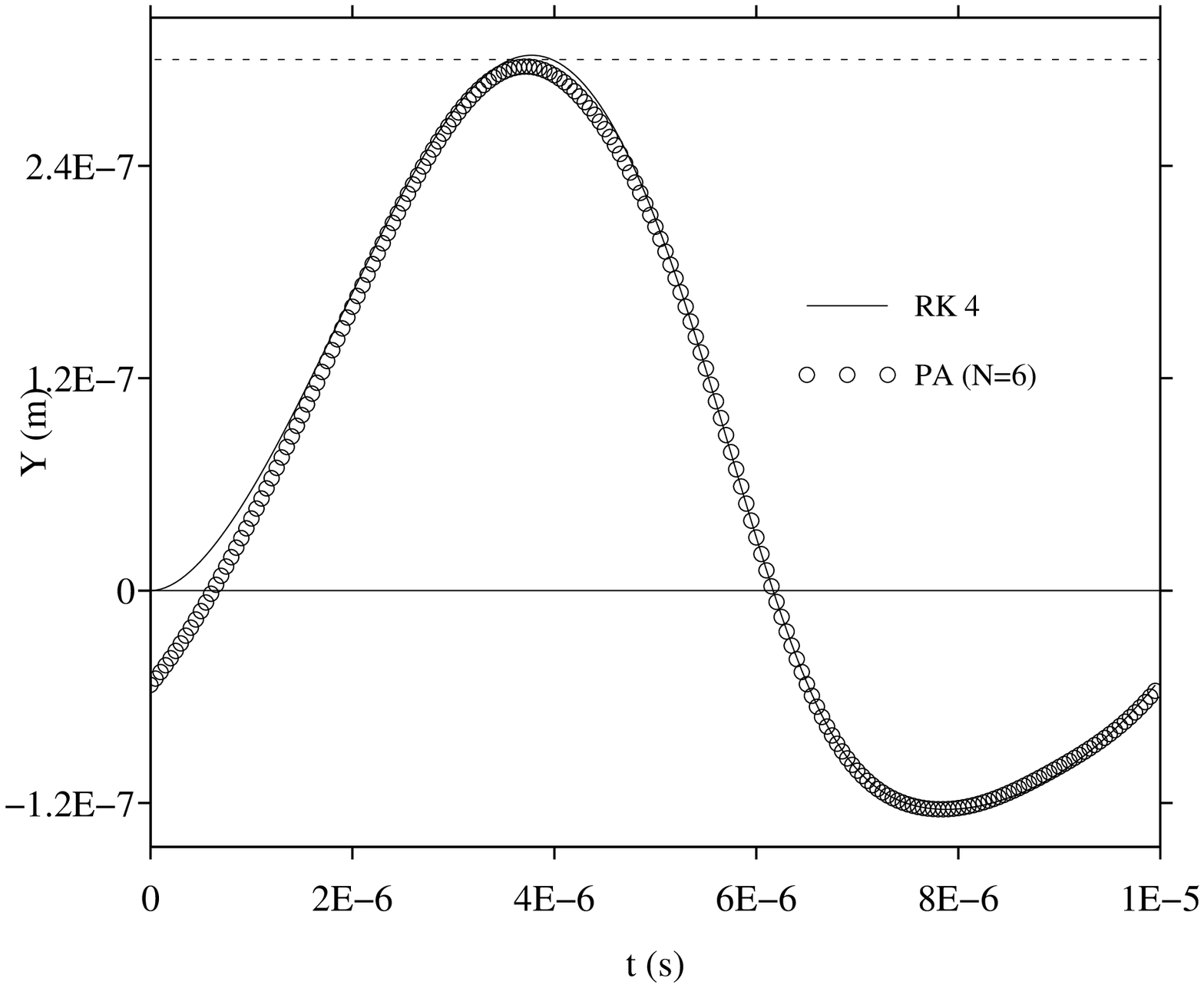}\\
(e) & (f)\\
\includegraphics[scale=0.28]{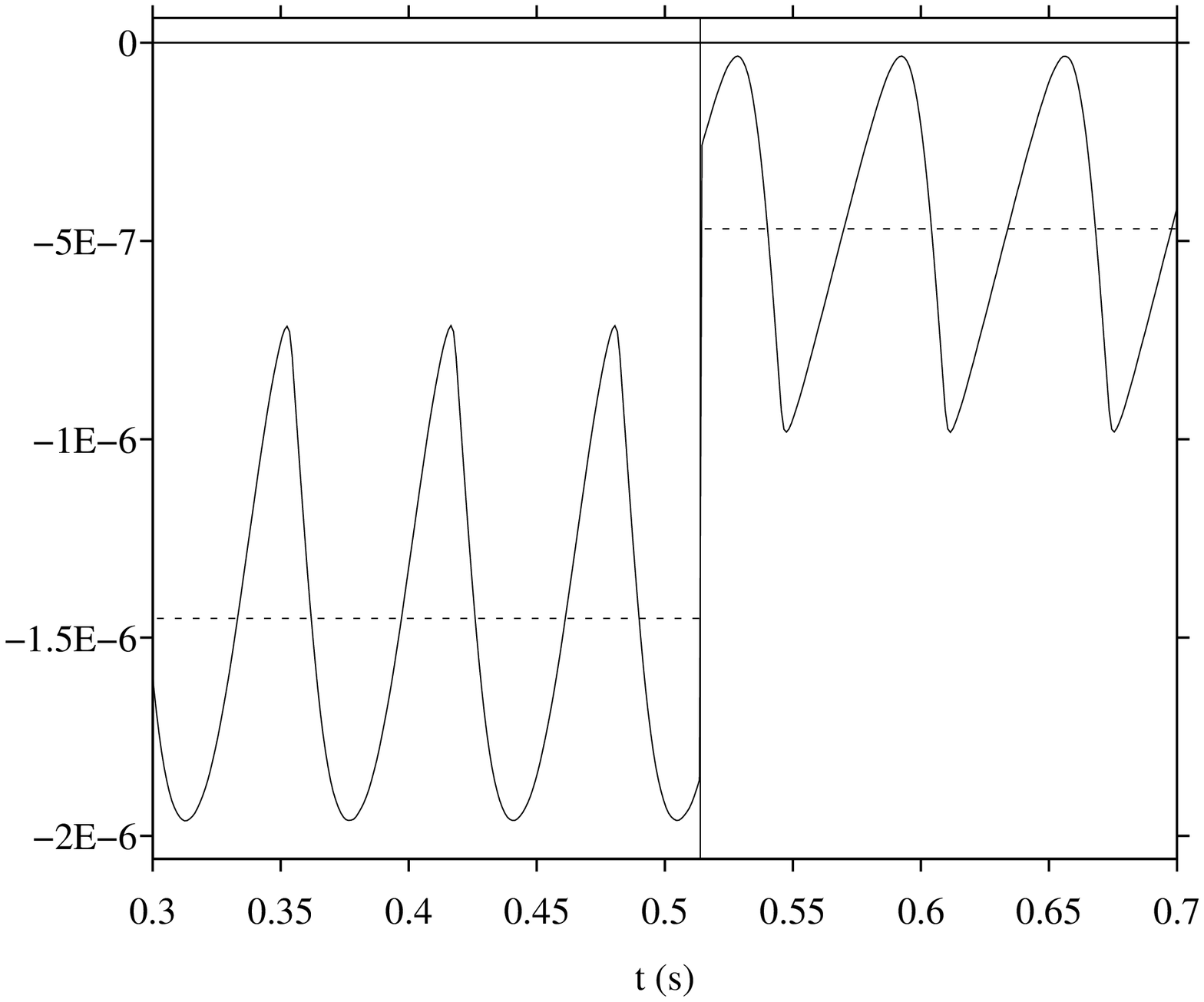}&
\includegraphics[scale=0.28]{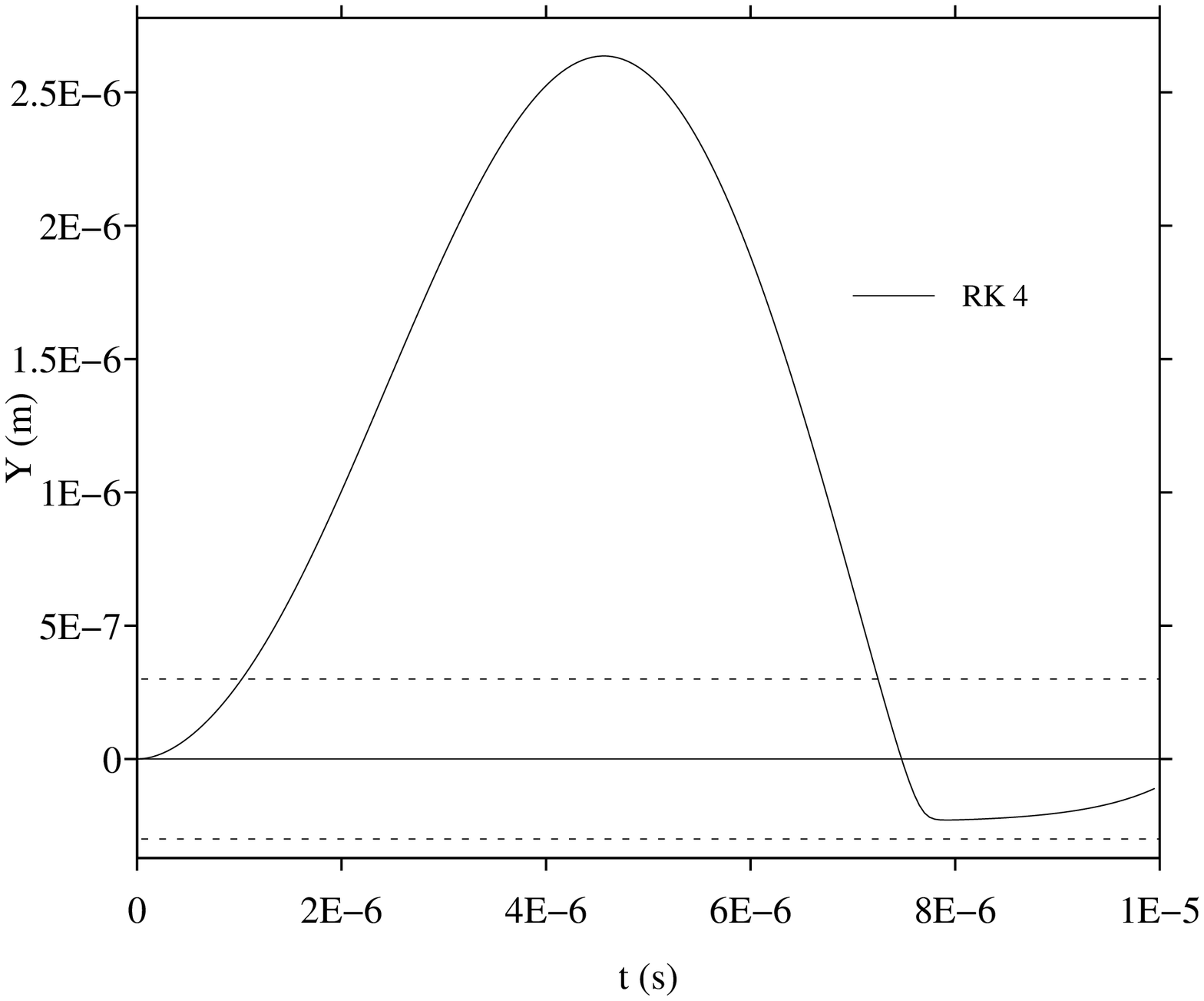}
\end{tabular}
\end{center}
\caption{Left column: snapshots of $u$ around $\alpha$ (the dotted horizontal lines denote the mean value of $u$, on both sides of $\alpha$). Right column: time history of $Y=[u(\alpha,\,t)]$ (the horizontal dotted lines denote $\pm \delta$). (a-b): $v_0=0.01$ m/s, (c-d): $v_0=0.13$ m/s, (e-f): $v_0=0.6$ m/s. }
\label{FigTest2}
\end{figure}

First, the elastic stress transmitted across the crack is shown in figure \ref{FigTest1}. In the left column, $\sigma_T$ computed by HBM ($\circ$) is compared with the DNS (-). In the right column, the normalized harmonics $\Gamma_i$ ($i=1,...,\,5)$ are shown. The amplitude $v_0=0.01$ m/s (a-b) is too small to mobilize the nonlinearity of the crack, and $N=2$ harmonics are sufficient. We measure $\Gamma_2=0.017$, to compare with $\gamma_2=0.016$: the approximation (\ref{PA_gamma}) is good. With $v_0=0.13$ m/s (c-d), $N=10$ harmonics are used, and $\Gamma_2=0.24$ is measured, to compare with $\gamma_2=0.25$. Lastly, with $v_0=0.6$ m/s, $N=50$ harmonics are required (e). We measure $\Gamma_2=0.60$, to compare with $\gamma_2=1.02$: the approximation (\ref{PA_gamma}) becomes poor. In the three cases and as deduced from the PA (section \ref{SubSecSingle}), the nonlinearity increases with $v_0$. At a given $v_0$, the $\Gamma_i$'s decrease strictly with $i$. 

\begin{figure}[htbp]
\begin{center}
\begin{tabular}{cc}
(a) & (b)\\
\includegraphics[scale=0.28]{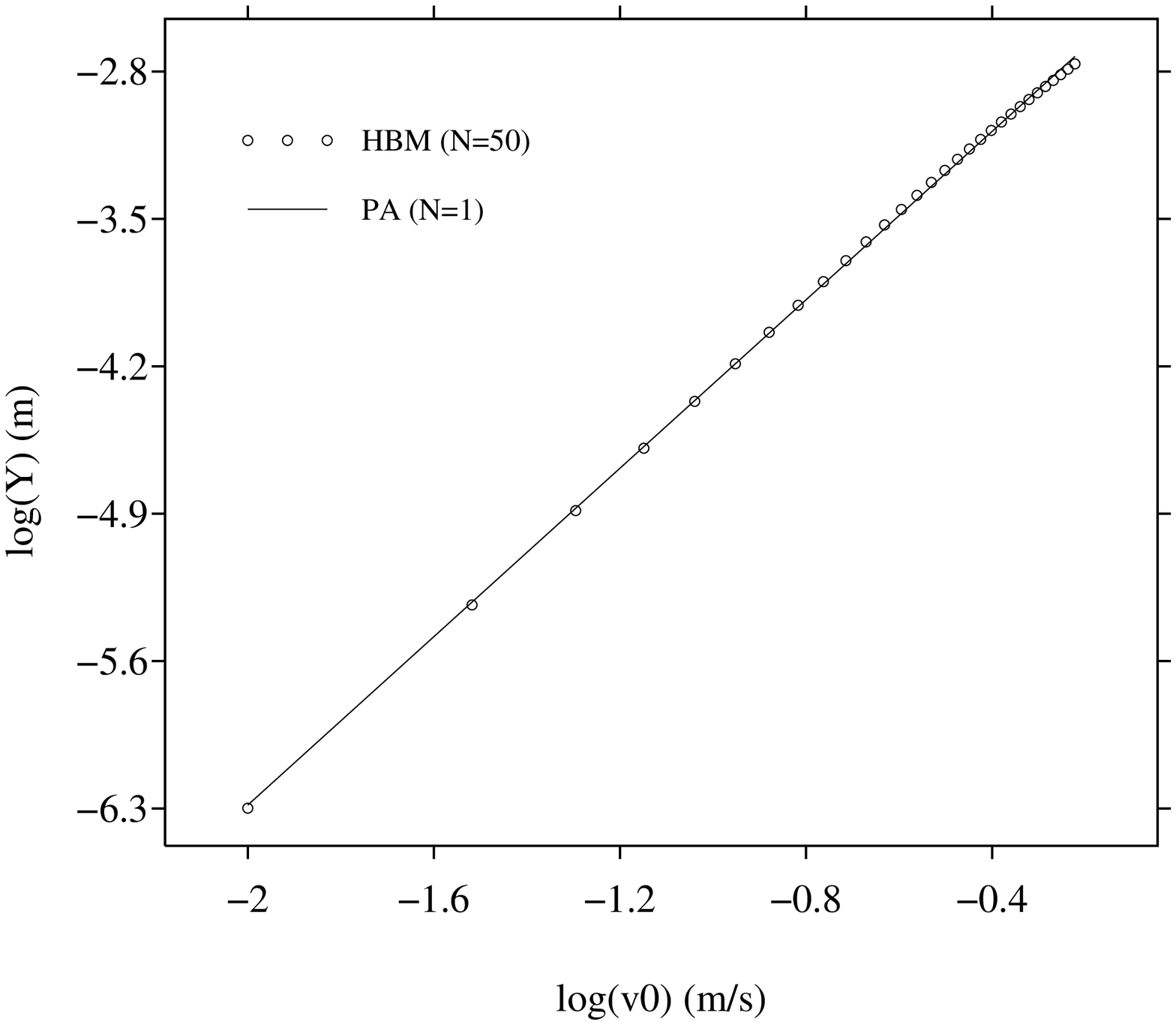}&
\includegraphics[scale=0.28]{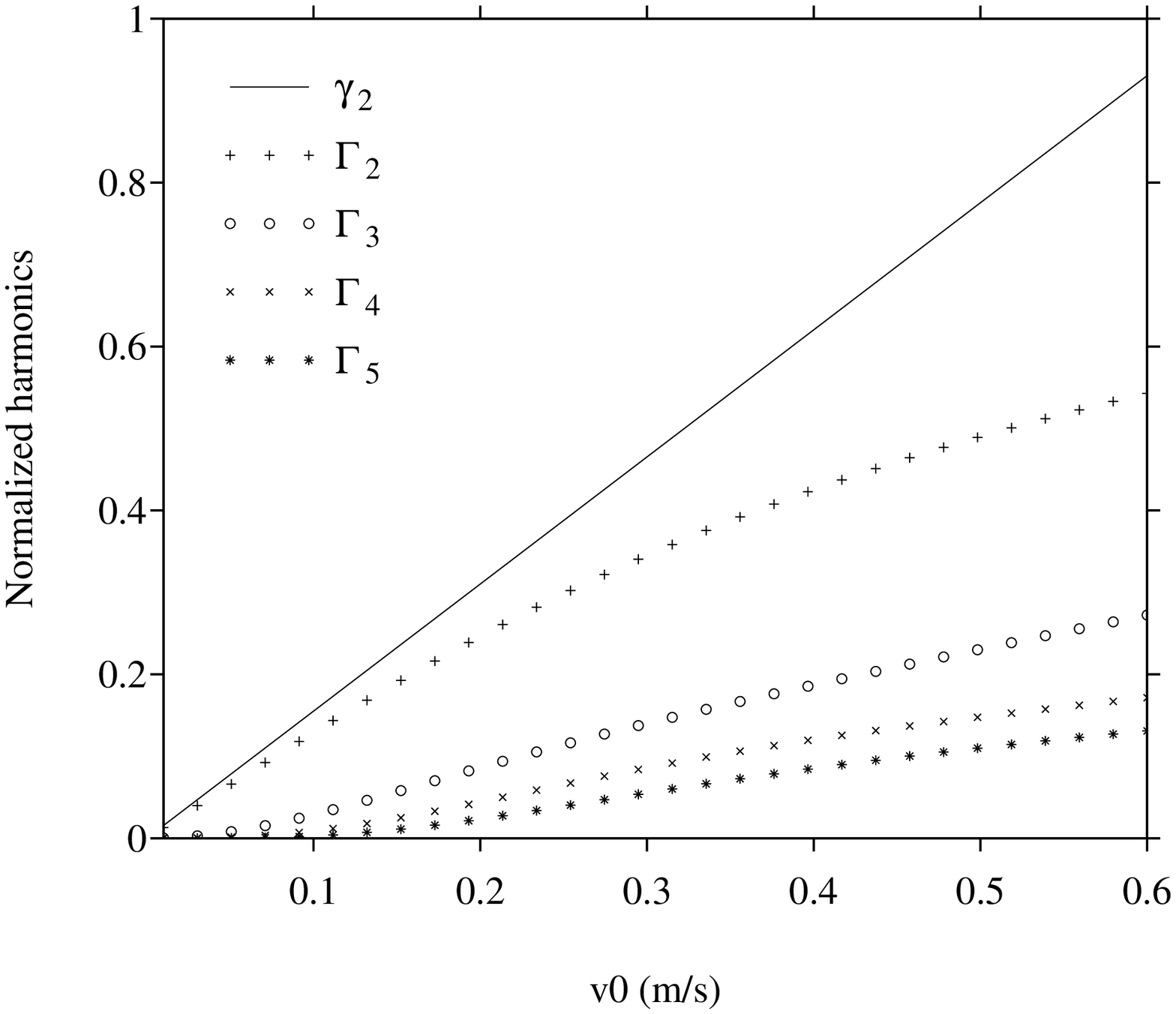}
\end{tabular}
\end{center}
\caption{Parametric studies in terms of $v_0$: $\overline{Y}$ (a), normalized harmonics $\Gamma_i$ ($i=2,...,\,5$) (b).} 
\label{FigTest3}
\end{figure}

Second, the influence of $v_0$ on the jump $\overline{Y}$ is illustrated in figure \ref{FigTest2}. The left column shows snapshots of $u$ computed by DNS. The jump between mean values of $u$ yields $\overline{Y}$. In the right column, the time history of $Y(t)=[u(\alpha,\,t)]$ is computed by fourth-order Runge-Kutta integration of (\ref{PA_ODE_Y}) (-) and by PA ($\circ$). At small $t$, PA differs from RK 4 because it does not compute transients. With $v_0=0.01$ m/s, $\overline{Y}=5.42\,10^{-11}$ m/s (to compare with the approximation $\overline{Y}_1=5.026\,10^{-11}$ m/s) is not visible (a), and $N=2$ in the PA (b). With $v_0=0.13$ m/s, $\overline{Y}=5.84\,10^{-8}$ m/s is measured (c), to compare with $\overline{Y}_1=5.60\,10^{-8}$ m/s, and $N=6$ is required (d). In that case, the maximum value of $Y(t)$ is roughly $0.99\,\delta$: with greater values of $v_0$, then $\max|Y|/\delta>1$, and the PA may not converge. It is what happens with $v_0=0.6$ m/s: only RK 4 is shown (f), and $\overline{Y}=9.82\,10^{-7}$ m/s is measured, to be compared with $\overline{Y}_1=1.02\,10^{-7}$ m/s. 

Parametric studies performed by the HBM ($N=50$) are proposed in figure \ref{FigTest3}. A log-log scale shows that $\overline{Y}$ is very close to the line with slope 2 deduced from (\ref{PA_DC1}), even at high $v_0$ (a). The amplitude of $\Gamma_i$ increases strictly with $v_0$, and $\Gamma_2$ tends towards $\gamma_2$ at small $v_0$ (b). 

\begin{figure}[htbp]
\begin{center}
\begin{tabular}{ccc}
(a) & (b)\\
\includegraphics[scale=0.28]{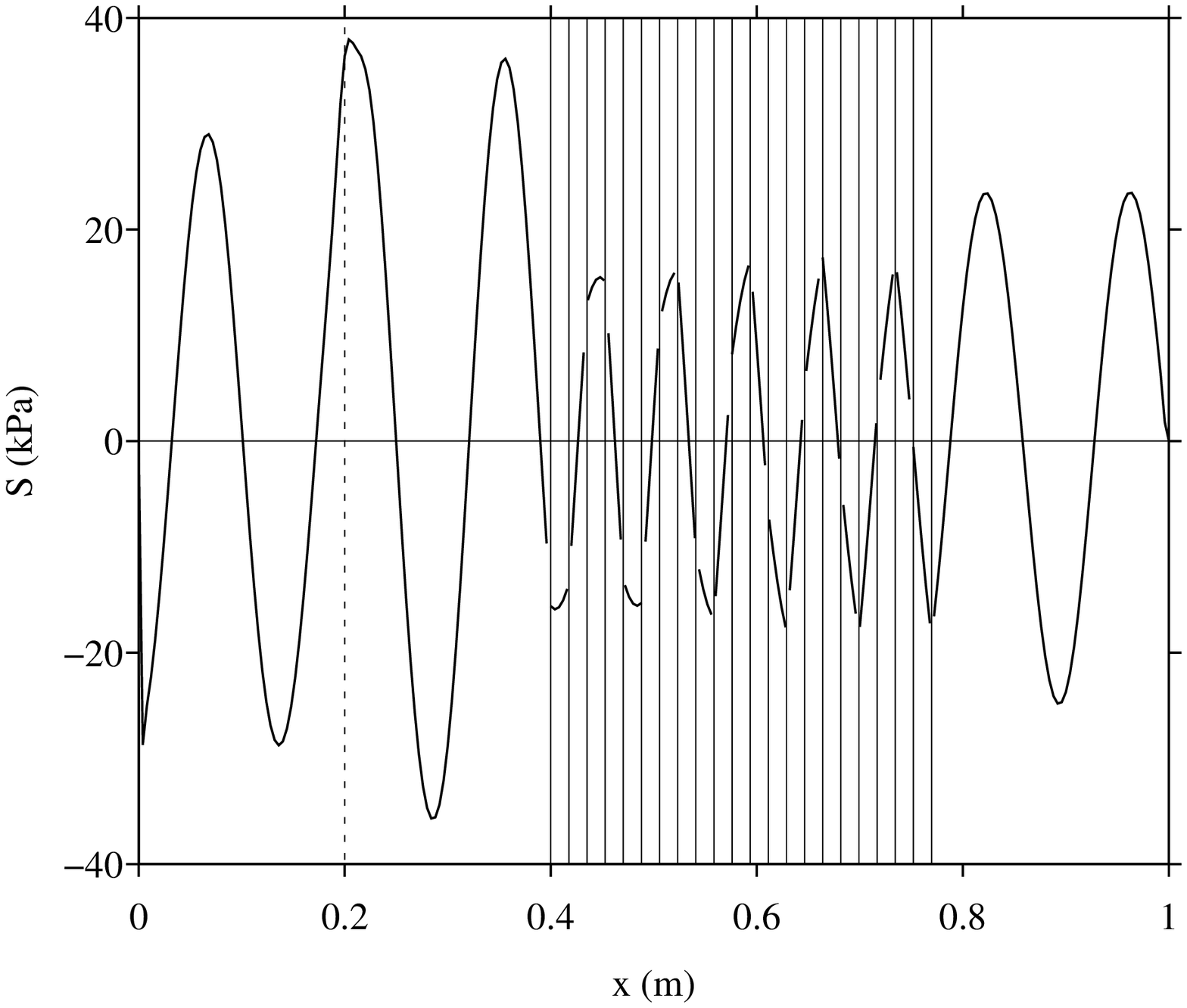}&
\includegraphics[scale=0.28]{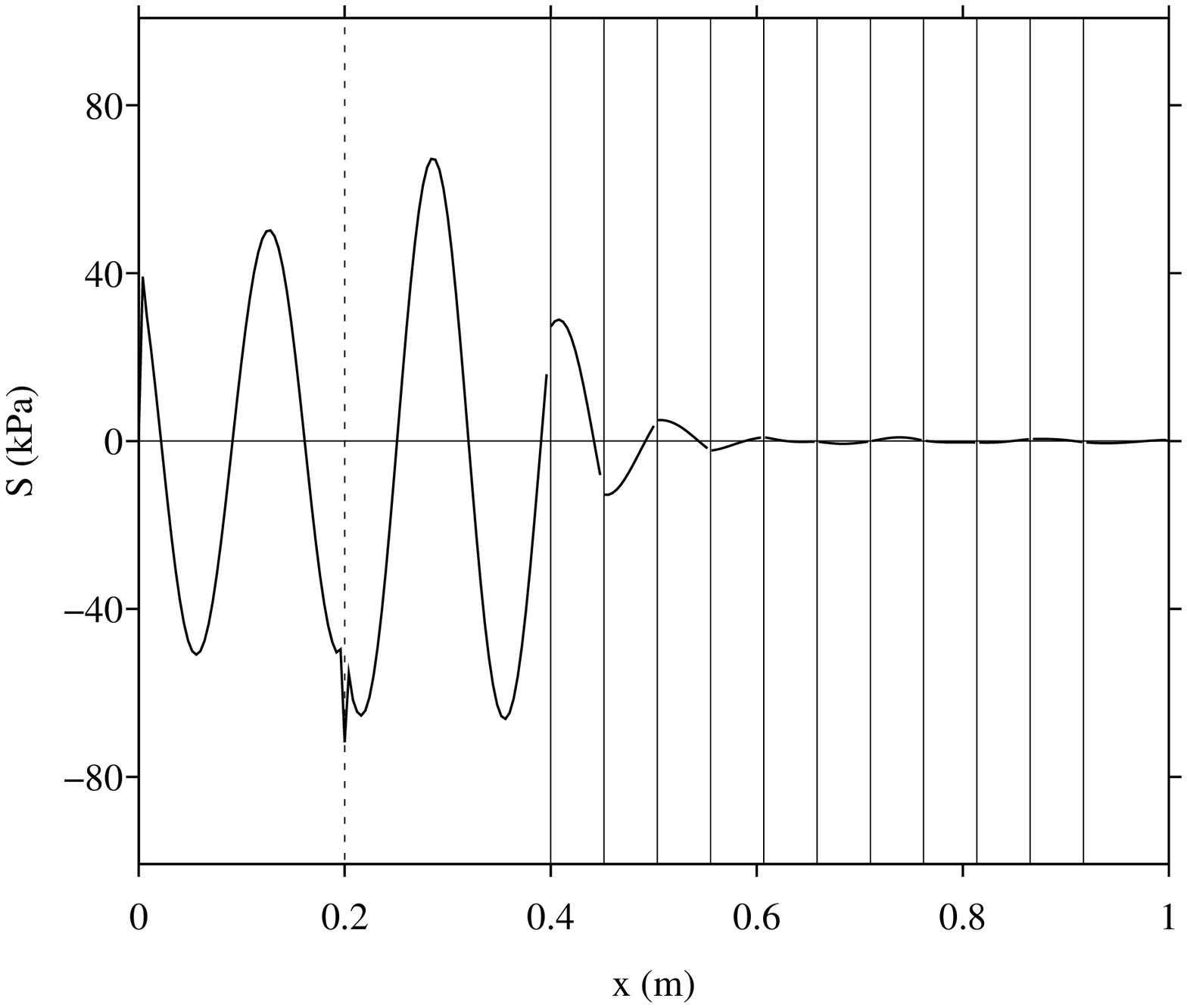}\\
(c) & (d)\\
\includegraphics[scale=0.28]{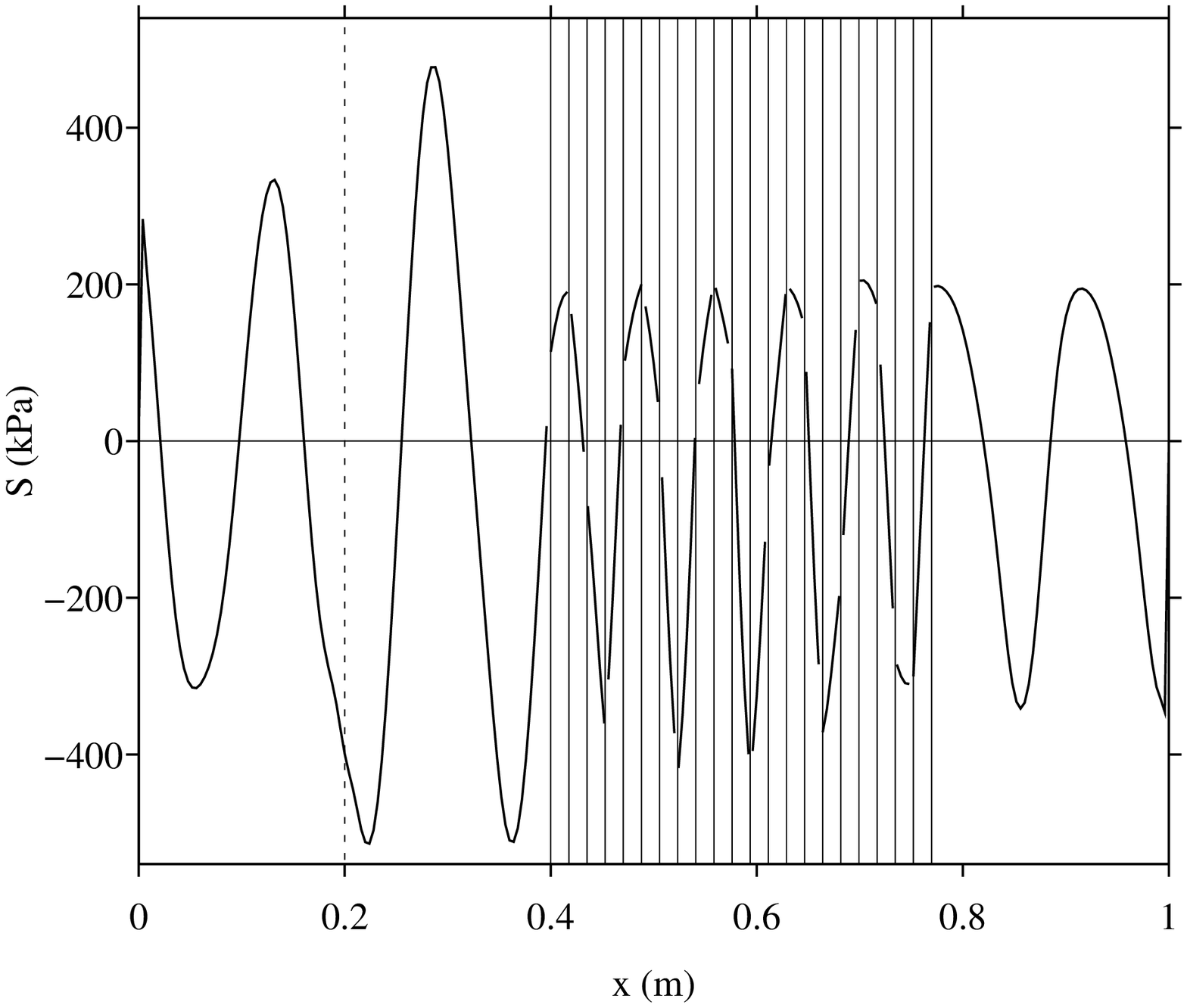}&
\includegraphics[scale=0.28]{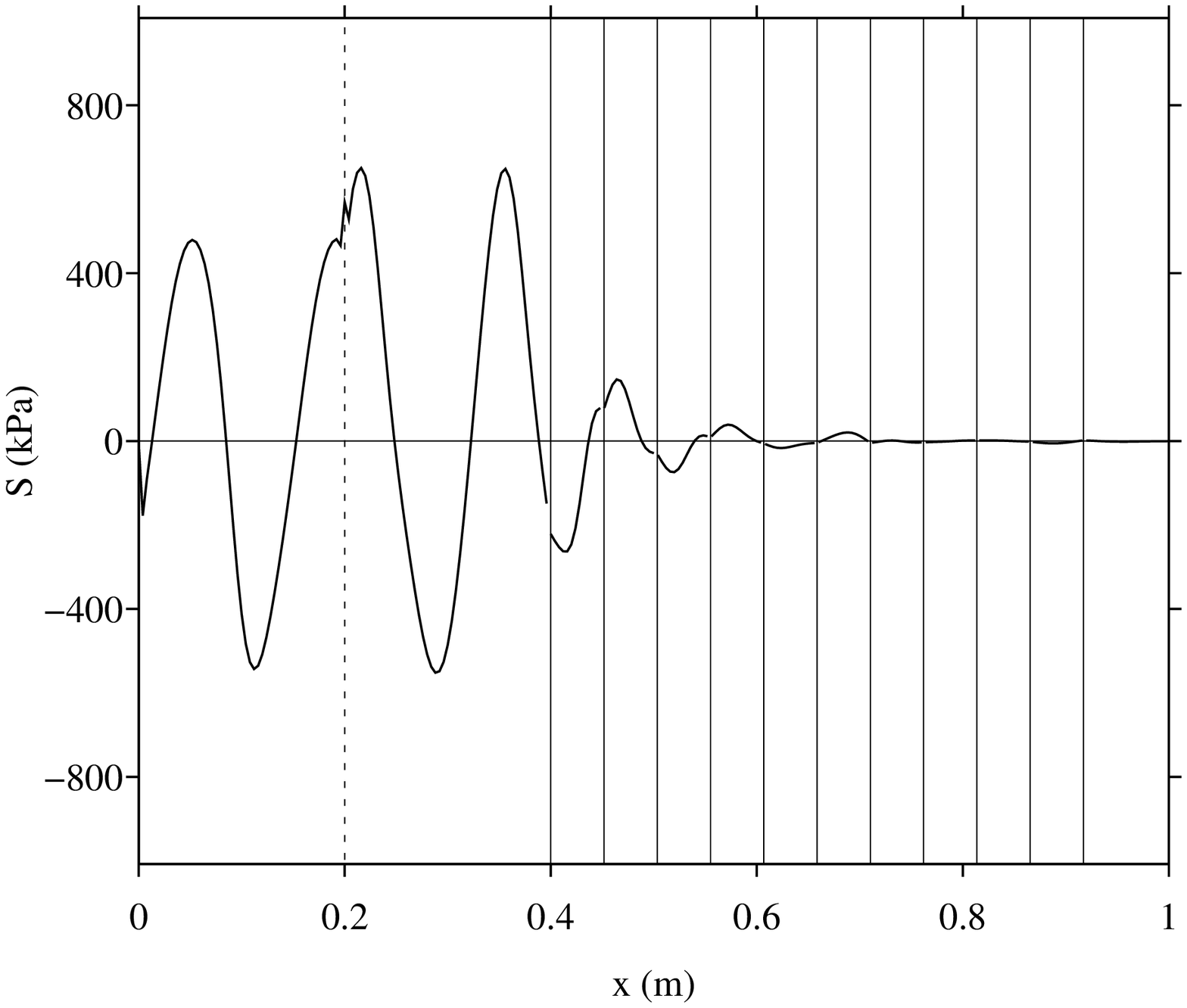}\\
(e) & (f)\\
\includegraphics[scale=0.28]{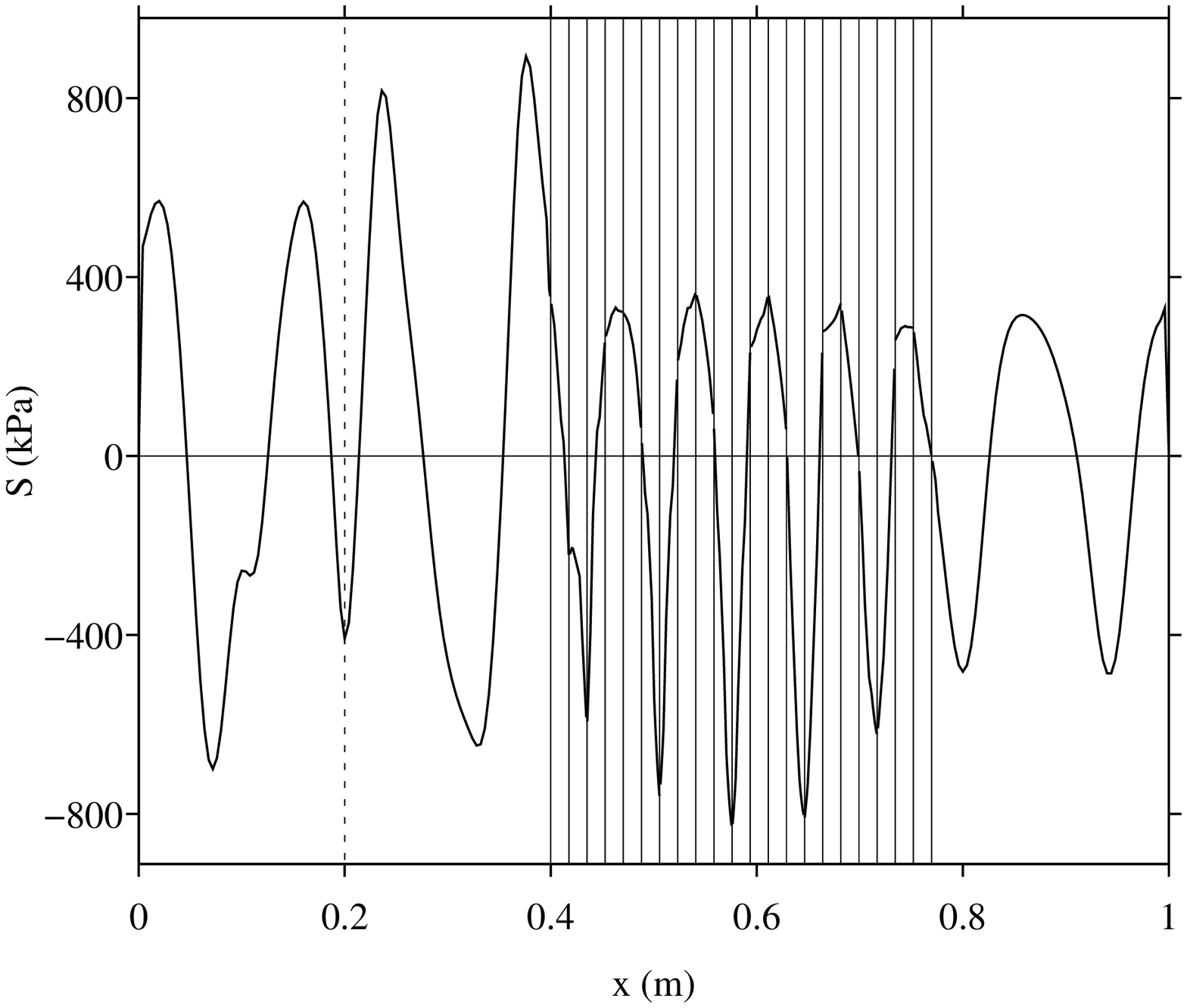}&
\includegraphics[scale=0.28]{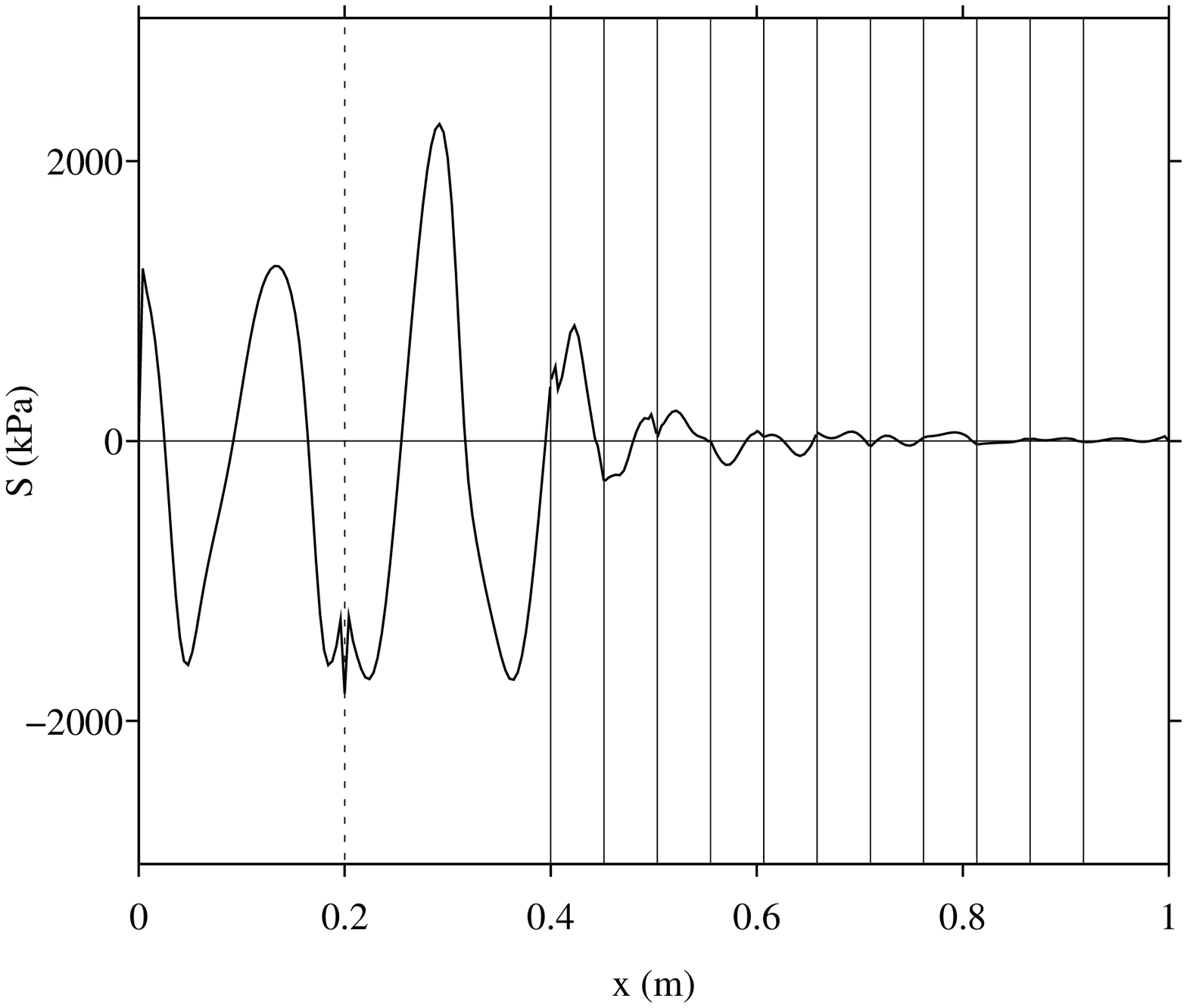}
\end{tabular}
\end{center}
\caption{Snapshots of $\sigma$, with $v_0=0.01$ m/s (a-b), $v_0=0.13$ m/s (c-d), $v_0=0.6$ m/s (e-f). The single vertical dashed line on the left denotes the source. The vertical solid lines denote the cracks.}
\label{FigTest4}
\end{figure}

Lastly, propagation across a periodic and finite network of cracks is simulated in figure \ref{FigTest4}. Two spacings are considered: $h=1.72\,10^{-2}$ m and $h=5.17\,10^{-2}$ m. In linear regime and with an infinite network, the Bloch-Floquet's analysis predicts respectively a pass-band and a stop-band behavior, observed with small amplitudes (a-b). The attenuation measured in (a) is in good agreement with the theoretical attenuation (\ref{FloquetAtt}). With higher amplitudes (c to f), the behaviors are maintained.


\section{Conclusion}\label{SecConclusion}

The main results of this work are as follows:
\begin{enumerate}
\item with one crack, two phenomena are induced by the nonlinearity: mean dilatation of the crack, generation of harmonics, that both increase with $\frac{v_0}{\beta\,\delta}$ and $\frac{\beta}{\omega}$;
\item with many cracks, simulations show that properties of infinite linear networks are valid in nonlinear regime with finite networks much greater than wavelength.
\end{enumerate}
Three directions are distinguished for further investigation:
\begin{enumerate}
\item investigation of shear effects, coupled or not with compressional efforts \cite{NORRIS82,PECORARI03};
\item analysis of the periodic solution of (\ref{PA_ODE_Y}), to prove $\overline{Y}>0$ and $\partial\,\overline{Y}/\partial\,\frac{v_0}{\beta\,\delta}>0$;
\item effective properties of a random linear and nonlinear networks \cite{BRADLEY94c}.
\end{enumerate}

\end{document}